# A classification
# of the natural and social distributions
# Part 2: the explanations

L. Benguigui and M.Marinov
Israel Institute of Technology
Solid State Institute
32000 Haifa
Israel

ABSTRACT

In this second part of our survey on the social and natural distributions, we investigate some models, which intend to explain the statistical regularity of the natural and social distributions. There is a large variety of models and in their majority, they look for a power law, at least in the tail, although there are several real distributions which are not described by a power law. Among the power law models, we discuss a) the two basic models and their variants: the random multiplicative model and the preferential attachment model; b) models based on the Bose-Einstein statistics; c) geographical, economical, and criticality models. We present also some models, which do not intend to explain a power law, and among them lognormal-like distributions, exponential and stretched exponential distributions. The interesting findings of this survey are that there are few models giving a power law for the complete distribution and that among them, the Zipf exponent 1 is rare.

# INTRODUCTION

In the first part of this work (Benguigui and Marinov (2015)), we conducted a survey of studies in natural and social sciences in order to determine, which functions are being used for the mathematical description of the distributions. Our survey consists of 89 cases in different disciplines and we presented the classification of the distributions into eight different types. The three representations are presented in the Appendix. One important finding is that the studies where the complete distribution is described by a power law (Strong Power Law) represent only around 30% of the cases. Sometimes, only the upper tail behaves like a power law, but the complete distribution is better described by another function. To be more precise we can say that there are three cases of "power law": A) The complete distribution is a power law (Strong Power Law); B) The distribution is well described by two different functions: a true power law in the tail, but the small items are described by another function; C) The distribution is described by one function for all the items and the upper tail is well approximated by a power law. Case C) is the most common of the three.

In this second paper, we examine a number of models, which were proposed in order to explain the observed distributions. We shall not give details of the calculations referring to the original papers: we shall expose the hypothesis of the model and the results.

Generally, it appears that majority of these models were developed in expectation of the power law. Sometimes only part of the distribution is included to accommodate this expectation, thus the cases of pure power law are relatively few. The literature on the subject is enormous, not all existing models could be discussed here, and we apologize for the missing articles. We should mention that in the quoted papers there are many references to the different models, which weren't discusses in this survey.

Some power law models are attempting to be general enough to be applicable to a number of different disciplines. These usually rely on a small number of hypotheses and assumptions. Other models of power laws are constructed for a particular sphere of knowledge (geography, economy, linguistics, earthquakes, biology, etc) and cannot be used outside their domain. We also discuss models, which produce other distributions, such as lognormal or exponential functions and finally we conclude with a general discussion.

# MODELS OF POWER LAWS

We begin with two important remarks. The large majority of the power law models aim to explain only one thing: how a power law emerges. They do not intend to give a general description of a phenomenon. For example, the original Simon model, which is discusses below, gives a linear increase in the number of cities with time, when applied to city size distribution. In reality it is not so: in the USA it is exponential and in Israel - quadratic (Benguigui and Blumenfeld-Lieberthal, 2009). Recently, this point was discussed by Pumain (2012), who complains about models giving only a power law and nothing more. Another limitation of some models is that they look for a steady state, when real systems are very often not in a steady state (like city size distribution or income distribution). Nevertheless, these models are important in a conceptual point of view and they may provide a reservoir of ideas to be applied in different cases.

## A. The minimal power law models

We call "minimal" a model in which there are very few hypotheses and may be applied to several real situations precisely due to this minimum of hypotheses. The explicit goal of these models is to explain the ubiquity of the power laws observed in so many different contexts. We begin by the most general model since it does not refer to any specific case in real life, but only to the distribution.

### 1. The most general model
B.Corominas-Murtra and Ricard Solé (2010) presented the properties of a distribution through its Kolmogorov complexity and its equivalence with the Shannon entropy. They consider a stochastic distribution of identically distributed variables and they get an equation relating the Kolmogorov complexity K to the normalized Shannon entropy. When a distribution changes the number of variables (generally by growing), it is assumed to maintain its basic statistical properties independently of the size of the system. Furthermore, it is assumed that the basic equation is applicable also to a part of the system. These hypotheses are referred to as "scale invariance conditions". The solution of the basic equation is a power law with an exponent dependent of the size of the distribution. It is possible to say that the power law is not a surprising result, when scale invariance is introduced into the model. The less trivial result is that the exponent converges to -1, when the size of the system goes to infinity. For a limited system the exponent may be less than -1 or larger, but it remains approximately -1. This explains why the exponents found experimentally are so often close to -1.

### 2. The random multiplicative process
The model has a large number of variations, such that numerous papers presented various solutions. One considers a series of items with size $S_i(t)$ which vary with time and one considers the change in their size by time t such that $S_i$ at time t+1 is related to its value at t by

$$S_i(t+1) = A\, S_i(t) + B \qquad (1)$$

where A and B stochastic variables, (not necessarily dependent on t) with distributions U(A) and V(B). It is also possible to generalize this equation by taking all the items as components of a vector and A and B as random matrices (Champernowne (1953), Kesten (1973)). This random multiplicative process received the name of "Gibrat process".

From this basic equation, one has different possibilities. The first is the possibility that $B \equiv 0$, conserving only the term $AS_i(t)$ in the right hand of (1). The second possibility concerns the case, where the number N of items is constant and one looks for a steady state adding some limit conditions, such as, for example: none of the $S_i$ may be less that a given value. The third possibility is varying the number of items by "death" or by "birth" or both. One can imagine other possibilities, but the three indicated here are the most widely applied in several variations of the basic model. Some models use only one possibility, while other models adopt several.

We shall begin with the simplest cases when $B \equiv 0$ and N is constant. It is a well-known result that the distribution of the items $S_i$ is a lognormal distribution. (Several papers give this solution, see Par and Suzuki (1973), Mitzenmacher (2003), Sornette and Cont (1997)). The question is: which conditions will generate a power law distribution?

*2.1 The number N of items is constant and $B \equiv 0$.*

Several articles were devoted to this case and some look for a steady state. The first idea was to impose a limit such that no item could be lower that a given threshold $S_0$ (Gabaix (1999, 2009), Levy and Solomon (1996), Sornette and Cont (1997)). When an item reaches a value less than $S_0$, it keeps this value until it begins to grow again. Two different possibilities were proposed to define this lower limit. The first is to take a constant value for this minimum size $S_0$. The second is to take the ratio $c = S_0/<S>$ as a constant when $<S>$ is the mean value of the item sizes.

In the first case, one gets a steady state with a tail following a power law if the mean value of (Ln A) is negative, $(<\ln A>) < 0$ (Sornette and Cont (1997). It is possible to write (1) as

$$\ln S_i(t+1) = \ln S_i(t) + \ln A \qquad (2)$$

and the quantity LnS varies as a random walk with step (Ln A). If $(<\ln A>)$ is negative the quantity $<S>$ will decrease and drifts toward $-\infty$. However, the introduction of a threshold for S gives the possibility of a steady state for $t \to \infty$. The threshold $S_0$ acts as a barrier which prevents S from going to $-\infty$. The PDF is given by

$$D(S) \sim S^{-1-\mu} \qquad (3)$$

where the exponent µ is solution of the equation

$$< A^{\mu} > = 1 \tag{4}$$

The exponent µ is not universal, it depends on the distribution U(A). Sornette and Cont (1997) showed that the exponent µ is given approximately by

$$\mu = \frac{(Mean\ Ln\ A)}{(Var\ Ln\ A)} = \frac{<Ln\ A>}{<(Ln\ A^2> - <Ln\ A>^2} \tag{5}$$

They take an example for which the distribution U(A) is uniformly distributed in the interval [0.48; 1.48]. In this case, the mean of (Ln A) is $< Ln\ A > = -0.006747$ and $< A > = 0.98$. The value of the exponent µ is 1.47. Finally, it is also possible to show that

$$\mu \approx \frac{1}{1-c} = \frac{1}{1-\frac{S_0}{<S>}} \tag{6}$$

It is concluded that in general µ > 1 and it converges toward 1 if $S_0 \to 0$.
In addition, there is a condition relating c and N, as was explained by Malcai, Biham and Solomon (1999). They show that the ratio $c = S_0/<S>$ and the number N of items must obey the following inequality

$$\frac{1}{Ln\ N} \ll c < 1 \tag{7}$$

The inequality (7) can be interpreted as a condition relating N, $S_0$, the mean and the variance of the distribution of A (the first two moments of U(A)). In case of $N \to \infty$, the inequality (7) is automatically satisfied.
Sornette and Cont consider also the case $(< Ln\ A >) > 0$ and one has to choose a threshold $S_1$, to keep S below or equal to $S_1$. With the minimal threshold $S_0$, the distribution recovers the lognormal distribution and the threshold is not important. With the upper threshold $S_1$ the distribution is again a power law for $S < S_1$, without getting a steady state since $< S >$ is an increasing function of time. Now the PDF is given by

$$D(S) \sim S^{-\mu-1} \tag{8}$$

for $0 < S \leq S_1$. The interesting point is that one can have two different regimes with $\mu > 1$ or $\mu < 1$. In the first case the PDF is an increasing function of S, beginning with S = 0 and in the second case, the PDF is a decreasing function with a divergence when $S \to 0$.

Sornette and Cont think that the case $(< Ln\ A >) > 0$ is irrelevant, but this is not evident if one wants to look for a complete distribution. It is possible to imagine a set of items divided in two groups with the stochastic variable A dependent of S. For the small values of S one has $(< Ln\ A >) > 0$ and $(< Ln\ A >) < 0$ for large values of S. One can also take the two thresholds equal ($S_0 = S_1$). One obtains two possibilities for the complete distribution depending if $\mu < 1$ or $\mu > 1$. In the first, for $S < S_0$ the PDF diverges as a power law for $S \to 0$ and decreases as power law with a different exponent for $S > S_0$. In the second, for $S < S_0$, the PDF decreases to zero as a power law and for $S > S_0$ it decreases as a power law for $S \to \infty$. In this second possibility, the complete distribution exhibits a maximum. These two cases may be called "double Pareto" distributions, as suggested by Reed (2002).

We may now consider the choice of the lower threshold $S_0$ (Malcai et al., 1999). It is not its value which is taken constant during growth, but the ratio $c = S_0/< S >$. This modification generates a power law, but for the complete distribution and not only for the tail as explained above. Thus we get the following:

A) A power law is observed for the whole distribution.
B) The exponent $\mu$ is independent of the distribution U(A), but is a complicated function of N and c.
C) The system may grow indefinitely and the only limit is given by the growth time. This means that $< A >$ is positive and consequently the condition $(< Ln\ A >) < 0$ is not fulfilled.

The exponent $\mu$ of the PDF (3) is strongly dependent on c but not too much on N. It increases with c from 0, which is equivalent to the lognormal distribution. The authors propose two approximate expressions for $\mu$ depending on c and N. If one has

$$\mu \approx \frac{1}{1-c} \tag{9}$$

and the exponent $\mu$ is larger than 1. But if $c \ll 1/N$, $\mu$ is given by

$$\mu \approx \frac{Ln\ N}{Ln\ (\frac{N}{c})} \tag{10}$$

and the exponent is smaller than 1.

From their numerical simulations, taking a uniform distribution between 0.9 and 1.1, they got for N=1000 and c = 0.3 a power law from $S = 10^{-4}$ to S = 0.3 with $\mu = 1.4$.

## 2.2 *The number N of items is constant with B ≠ 0*

It exits two versions of the equation (1), the discrete version given by (1), when the time changes by steps and the continuous version in a form a differential equation. Equation (1) may be written as

$$S_i(t + 1) - S_i(t) = (A - 1)S_i(t) + B \qquad (11)$$

Passing to the continuos limit $S_i(t + 1) - S_i(t) = dS$ gives

$$dS = CS(t)dt + \sigma S(t)\, dW \qquad (12)$$

W(t) may be seen as a noise added to the evolution of S.

Several authors (Kesten (1973) Gabaix (2009)) give the solution of (1) and here we present that of Takayasu et al (1997) (with B > 0). They got the following results: the power law distribution of the tail is given as above by (3) and the exponent µ is the solution of the same equation (4). They add to the condition (<Ln A>) < 0, the condition $(< A^2 >) > 1$. The term B acts as a positive force which prevents S from going to −∞ similar to the minimal threshold discussed above. This is why the exponent µ is determined through the same equation (4) and B appears only in the coefficient of the PDF.

The interesting point in this work is the calculation of the complete CDF, while choosing explicit forms for the distributions U(A) and V(B). U(A) is Poisson distribution taking discrete values for A = 0, c, 2c, 3c, … and V(B) is a Gaussian distribution. The CDF exhibits effectively a power law behavior for large S with exponent µ, positively dependent on parameter c. However, for S → 0, the CDF goes to 1 with an apparently null slope. The PDF D(S) may have a maximum and go to zero with S as a power law, generating a double Pareto distribution.

Gabaix (2009) proposes an interesting model of birth and death with an assumption of constant number of items. An item dies in a Poisson process with rate δ and is reborn immediately with a new size $S^*$. There is no minimum threshold or reflecting barrier. Solving the continuous equation (12) gives the following results. The exponent of the distribution is the solution of a second-degree equation with two solutions $µ_1 < 0$ and $µ_2 > 0$:

$$(\sigma^2/2)\, µ^2 + (C - \sigma^2/2)µ - \delta = 0 \qquad (13)$$

Thus the PDF of the distribution has two parts:

$$D(S) \sim S^{(-\mu_1 -1)} \qquad S < S^* \qquad (14a)$$
$$D(S) \sim S^{(-\mu_2 -1)} \qquad S > S^* \qquad (14b)$$

The absolute value of $\mu_1$ may be larger or smaller than 1 depending of the rate $\delta$. If it is large enough, the quantity $(-\mu_1 -1)$ is positive and $D(S < S^*)$ goes to zero for $S \to 0$. But if it is small enough, the quantity $(-\mu_1 -1)$ is negative and $D(S < S^*)$ diverges when $S \to 0$. This again results in the double Pareto distribution, since for $S > S^*$ in the tail is a power law.

*2.3 The number of items is not constant and $B \equiv 0$*

These models necessitate the introduction of mechanisms for the apparition of new items (birth process) and their eventual disappearance (death process). They do not give a steady state, only a solution after a time long enough to establish stability. We will present only two out of the many examples of such models.

Blank and Salomon (2000) use the equation (1) and introduce the two processes as follows. An item will disappear if its size goes below a given threshold $S_0$. The threshold defines the minimum value of S and consequently the equation (7) must be satisfied. At each step in the process, a number $\Delta N$ of items are introduced proportionally to increase the total number of items:

$$\Delta N = K\,[N(t+1) - N(t)] \qquad (15)$$

where K is a constant. The regular increase of N allows for (7) to be satisfied. With this choice of $\Delta N$ the ratio $c = S_0/<S>$ is a constant equal to $KS_0$. Blank and Solomon make a numerical simulation with $<A> = 1.02$ and $(Var\ A) = 0.0025$ and they get a power law for the complete distribution with an exponent slightly larger than 1, given again by

$$\mu \approx \frac{1}{1-c} = \frac{1}{1-\frac{S_0}{<S>}} \qquad (16)$$

Blank and Solomon apply their model to cities, income and firms.

Benguigui and Blumenthal-Lieberthal (2007) created a numerical simulation analogical to the model by Blank and Solomon with application to cities. Instead of the equation (13), they propose the introduction of one new item (with the minimum size below which an item disappears) after K steps in the evolution of the system. The interesting aspect of their work is the fact that the distribution is determined as a function of time and consequently one can follow the evolution of the distribution. Parameter K may be a constant or a function of time through the steps of the growth process. As it turns out, the resulting distributions may belong to type A (A1, A2 or A3), to type B or to type C.

If K is large, for short times the distribution is of type A and it becomes type C for longer times. If K is small enough, the distribution passes through the three types: for short times it is of type A, for intermediate times - of type C and for long times it is of type B. Let us recall, that type A may exhibit a power law behavior in the tail or no power law at all, for type C the upper tail is power law and type B is a complete power law. The exponent µ is a strongly increasing function of K, but it is independent of time, once the system enters the region of type B. For K = 50 the exponent is 0.5 and increases until 1.5 for K = 200.

*2.4 The number of items is not constant and B ≠ 0.*

The model by Malevergne et al (2008) is developed for the distribution of firms, but it can be applied to other situations with minor changes. The basic equation is the continuous version (12) with C as a constant. The authirs introduce mechanisms of birth and death. The model has four hypotheses:

1. There is creation of new items as a random flow, following a Poisson distribution with parameter d.
2. The initial size of the new items is given by an exponential increase with parameter $c_0$.
3. A firm disappears if its size goes below a given threshold, which changes with time at a rate $c_1$ such that $0 < c_1 < c_0$.
4. Firms may disappear independently of their ages and their sizes at a constant rate h. This hypothesis corresponds to the case of firms disappearing due to economic shock. When applying the model to city size distribution, h may be set to equal zero.

The tail of the distribution is a power law with exponent µ dependent on the parameters C and σ (of the equation (12)), $c_0$, d and h, for times larger than a limit also dependent on these five parameters.

The most interesting result is the variation of µ with the ratio $k = (d + h)/(C - c_0)$. If k <1, the exponent is also smaller than 1. It becomes equal to 1 for k = 1 and larger than 1 for k > 1. Malevergne et al. (2008) discuss these results in the context of distribution of firms in an economy.

We would like to note an important and rare analysis in the work of Malevergne and the coauthors. The calculations give the mean of the distribution. However, a real world distribution, due to its stochastic character, corresponds to one realization. By means of numerical simulation, the authors check that the realization of their model is not far from the calculated mean. This is an important lesson about models based on numerical calculations: there is no need to calculate the mean of several realizations of the distribution in order to compare a model to a real world distribution. It is enough that at least one realization of the model is in agreement with reality, to say that the model is good.

Reed in several publications (2002, 2003) developed a model of growth, which he applied to cities and incomes. The model includes only birth, but no death. His starting point is the equation (12) with some additional assumptions. The first concerns the size $S_1$ of the new items, which are distributed following a lognormal distribution with mean $C_1$ and variance $\sigma_1^2$. The second assumption concerns the ages of the items. At a time t, the distribution of the ages (times from their apparition until time t) is a decreasing exponential distribution with parameter $\lambda$.

The exponents of the S distribution are the solution of a second order equation analogous to the equation (13) when $\delta$ is replaced by $\lambda$. The final results are formally identical with the equations (14a and 14b) for $S \to 0$ and $S \to \infty$

$$D(S) \sim S^{(-\mu_1 - 1)} \quad (S \to 0) \tag{17a}$$
$$D(S) \sim S^{(-\mu_2 - 1)} \quad (S \to \infty) \tag{17b}$$

The properties for the distribution are the same: for $S \to \infty$ a tail with exponent $\mu_2$ and for $S \to 0$, two possibilities, divergence or decrease to 0. The mathematical expression of the complete distribution is very complicated and one can be satisfied by the two limits (14a) and (14b).

As a final note, one can remark that Reed calls his model "Double Pareto Lognormal" (DPLN), because the introduction of a lognormal distribution in one the hypotheses of his model. He applied his model to cities and income size distributions.

*2.5 Some remarks on the Gibrat model*

The following table gives a survey of the results discussed above.

|  | B = 0 | B ≠ 0 |
| --- | --- | --- |
| Constant N without constraint | Lognormal | Likely type A1 with power law in the tail Double Pareto (?) |
| Constant N with constraint | Power law in the tail or Complete Power law | Power law in the tail or Double Pareto |
| N variable and increasing | For long times Power law For short times type A or C | Power law in the tail or Double Pareto |

**Table 1: Results of the Gibrat model with different constraints**

One can note the following points. The original equation gives a lognormal distribution for a constant number of items. This distribution is characterized by the following properties: when S → 0, D(S) also goes to zero; for large S, there is a decline, which may be confused with a power law; in the mid-range the PDF exhibits a maximum.

The extensions of the different models are of two types: either they place a constraint in order to obtain a steady state different from the lognormal or the number of items increases with the growth of the item sizes. In all the cases, after some time (which was rarely defined) the tail is a power law with an exponent µ of CDF, which may be smaller or larger than 1. The case µ = 1 does not seem to be an exception, besides in the work of Malevergne et al (2008).

However, it is likely that the behavior of the PDF for the small values of S (for S → 0) is either a double Pareto law with a decrease to zero or a divergence. The divergence may be with the same exponent as for the large S and thus generates a complete power law (type B). In other cases, the divergence appears with a different exponent and the graph Log(Size) versus Log(Rank) is a compound of two straight lines (Type E). In other cases, the distribution has a maximum (same as in the lognormal), but now with a behavior of power law for the small sizes (type A1).

3. *The preferential attachment model*

The model and its different versions are very well known and were applied to several distributions. We begin with the model by Simon (1955), who calls it "the Yule distribution".

3.1. *The Simon model and its principal versions.*

Suppose a person is reading a book and at a certain moment, he has read k words. The number of different words, which appeared S times at step k is f(S,k). The next (k+1)th word is either a word, which was already read and has appeared S times and the probability of occurrence is proportional to Sf(S,k); or it is a new word, that has not appeared until this step and its probability to occur is α.

The PDF of the words frequency (after many words are ready such as k → ∞) is

$$D(S) \sim B(S, 1+\rho) \qquad (18)$$

where B(x, y) is the Beta function

$$B(x, y) = \frac{\Gamma(x)\Gamma(y)}{\Gamma(x+y)} \qquad (19)$$

and $\rho = 1/(1 - \alpha)$. For large x, the Beta function can be approximated by $B(x, y) \approx x^{-y}$. Thus for large S (the tail of the distribution) the PDf is a power law

$$D(S) \sim S^{-1-\rho} \qquad (20)$$

Since $\alpha < 1$ it results that $\rho > 1$. The possibility of a PDF exponent smaller than 2 is excluded in this model. Thus, for the Zipf representation the exponent is smaller than 1.

$$D(S) \sim S^{-1-\rho} \qquad (21)$$

Since $\alpha < 1$, it results that $\rho > 1$. The possibility of the PDF exponent smaller than 2 is excluded from this model. Thus, for the Zipf representation the exponent is smaller than 1.

Simon presents his model with the explicit intention to apply to different disciplines. He applied it to distribution of words, scientific publications, city sizes, incomes and biological species. For example, for the case of cities, the model describes the growth as follows. One has an infinite reservoir of inhabitants of the developing system of cities. The first inhabitant creates the first city. The successive inhabitants (one by one) come to the system with a choice: either to go to an existing city with probability proportional to its size (i.e., the number of inhabitants in this city) or create a new town with a constant probability $\alpha$.

Zanette and Montemuro (2005) proposed a new version of the Simon model in order to make it more flexible for comparison with the real data. They introduced two modifications: in the first, the introduction of new words is not made with a constant probability $\alpha$ but with a probability dependent of k as $\alpha_0 k^{v-1}$ with $0 < v < 1$. The probability that each introduced word will be a word, which already has occurred is not simply proportional to number of times it has already occurred, but to maximum between this number and a parameter $\delta$: max{f(S,k), $\delta$. The new parameter $\delta$ is a stochastic variable with an exponential distribution. With these modifications, Zanette and Montemuro succeeded in getting the exponent $\rho$ greater or smaller than 1 like in numerous real cases. At the same time, the modelled curves are in better agreement with some real distributions: words frequencies for English, Spanish and Latin.

Following Barabasi and Albert (1999), Clauset (2011) proposed a simplified version of the Simon model, applied to the development of a network. One begins with a network of two vertices or nodes joined by a single edge. At each step a new vertex is added and it joins an existing vertex i with a probability dependent on the degree $S_i$ of the chosen vertex:

$$P(S_i) = \left(\frac{r+S_i}{\Sigma_j(r+S_j)}\right) = \frac{r+S_i}{nr+n<S>} = \frac{r+S_i}{n(r+c)} \qquad (22)$$

In (18), r is a parameter and c is the mean degree of the network. Taking c as a constant of the network, one gets that the PDF of the network degrees is

$$D(S) = \frac{B(S+r, 2+\frac{r}{c})}{B(r, 1+\frac{r}{c})} \qquad (23)$$

where B(x,y) is again the Beta function. For large S, the PDF becomes a power law

$$D(S) \sim S^{-\gamma} \qquad (24)$$

with γ = 2 + r/c . In this case, the PDF exponent is larger than 2, same as in the original Simon model.

Krapivsky et al (2000) developed a similar mode. The probalilty of a new node to be connected to an existing node with S degrees is proportional to $S^\gamma$. If γ < 1, the number of sites with S links (or degrees) is a strectched exponental, but when γ > 1, a single site is connected to almost all the other sites. However, the case of γ = 1 corresponds to a power law with exponent larger than 2.

An interesting version of the model is the Chinese Restaurant Process (CRP), which became known to physicists only recently (Basseti et al. 2010). It is somewhat surprising, that this version of the model was not applied to a real-life distribution, such as, for example, city size distribution. The model was conceived as a Chinese restaurant with an infinite number of tables and an infinite set of customers. The first customer entering the restaurant occupies the first table and the other customers come in one by one. A new customer at a time T + 1 may either go to an already occupied table with S customers or occupy a new table. The probability to go to an already occupied table with $S_i$ customers is

$$P_i = \frac{S_i - \beta}{T + \theta} \qquad (25)$$

The probability to occupy a new table is

$$P_n = \frac{\beta N(T) + \theta}{T + \theta} \qquad (26)$$

where θ is positive parameter and $0 \leq \beta < 1$. N(T) is the number of occupied tables. The main difference from the Simon model is that the probability to occupy a new table is not constant but decreases with time (because N(T) grows slower than T).

The PDF of the number of customers per table is proportional to

$$D(S) \sim \frac{\Gamma(S - \beta)}{\Gamma(S+1)\Gamma(1 - \beta)} \qquad (27)$$

which becomes a power law for large S, $D(S) \sim S^{-1-\beta}$. In this version of the model, the PDF exponent can't be larger than 2 (more precisely, it is between 1 and 2), while in the Simon model it can't be less than 2.

To sum up these examples of the Simon model, we remark that the PDF exponent is closely related to the two probabilities of the model and can generate large variation of the PDF exponent. All the above-mentioned models belong to type C2, where a large portion of the PDF resembles the power law (see Appendix).

3.2 *The Simon model with birth and death*

Maruvka et al (2011) presented a version of the Simon model with birth and death of items. It is possible to view the original model as though it includes only birth of new items in the system at each step. The extension of the model is as follows. At each step, an item is chosen at random among all the items forming groups of $S_i$ items. With a probability 1- p, the item is removed from the system, i.e. it is dead. With a probability p it reproduces itself and a new item is added, i.e. it is born. The newly born item may be in the group in which it is born with probability 1 - m or it may create a new group with probability m. The growth rate of the system is γ = 2p -1. The behavior of the system depends on the relative magnitudes of m and γ. If γ > m, the asymptotic behavior of the PDF for large S is

$$D(S) \sim S^{-1-v} \qquad (28)$$

where v = m/(γ – m). But if m > γ, the asymptotic behavior of the PDF for large S is

$$D(S) \sim S^{-1-v} \qquad (29)$$

but now with $v = \gamma/(m - \gamma)$.

The complete expressions of D(S) are complicated, but from the curves describing some real cases, one can conclude that these theoretical distributions belong to type C2.

## **B. Models based on the Bose-Einstein statistics**

Yukalov and Sornette (2012) proposed a model with the intention to explain the apparition of the King effect or in their terminology the Dragon-Kings. In some distributions the largest item (that with rank 1) is so large that it cannot be included in the distribution together with other items. This is the King effect. For cities, this is the case with Paris relative to all the other cities in France. Another example are Moscow and Saint-Petersburg in Russia.

The model is based on an analogy with the Bose-Einstein condensation of bosons atoms. The bosons atoms are distributed in energy following the Bose-Einstein statistics, which allows two atoms to be in the same quantum state. When the temperature is below a characteristic level, the lowest energy state becomes more and more populated, such that this state cannot be included in the energy distribution of the atoms. The analogy is clear: the atoms are the inhabitants, the energy levels are the ranks, and the occupation of an energy level is the number of inhabitants. The reason for the condensation, which is the atom interaction, is replaced by competitive interaction between cities. The condensation itself is the apparition of the King city or the Dragon-King. The model is based on the following assumptions:

1. A utility factor $\omega$ can be defined for each city, which expresses its attraction strength.
2. This utility factor is an exponential function of the rank R of the city

$$\omega(R) = b \exp(-\beta R) \qquad (30)$$

b and $\beta$ are positive constants.

3. The probability, that a city of rank R has at least S inhabitants is given by

$$p_S(R) = a(R)\omega^S(R) \qquad (31)$$

The first step is to calculate the function a(R) by summing the probabilities and to equate the sum to 1. One gets the following expression

$$a(R) = 1/(1 - \omega(R)) \qquad (32)$$

And the size S of a city is given by

$$S(R) = \sum S\, p_S(R) \qquad (33)$$

where the sum is over all the sizes. Taking into account the equations (39) and (40) one gets for the function S(R)

$$S(R) = \frac{e^{\beta(R-\mu)}}{e^{e^{\beta(R-\mu)}} - 1} \qquad (34)$$

with μ = (1/β)Ln b.

This is the exact expression of the probability of occupancy of level R following the Bose-Einstein statistics. The parameter μ is equivalent of the chemical potential, which has an important role in the apparition of the Bose-Einstein condensation.

At this stage, the authors add a new limitation on the rank. If C is the total number of cities, the rank of the last city is C and as the minimal number of inhabitants in the last city is defined as m. Thus, R(m) = C. This expression together with (34) gives for the chemical potential μ

$$\mu = C - \frac{1}{\beta} \text{Ln}\left(\frac{m}{m-1}\right) \qquad (35)$$

In the Bose-Einstein statistics, the chemical potential is smaller than the lowest energy, which is rank 1, hence $\mu < 1$. When the temperature drops, the chemical potential increases until it reaches value 1, and the condensation begins. The lowest state becomes more and more populated, when the chemical potential remains practically equal to 1.

Supposing that C and m are constant and that parameter β (analogous to inverse of the temperature) increases, the chemical potential may reach critical value μ = 1, such that the King effect appears. A large value of β implies a stronger attraction for the small values of R, than for the larger. This explains the apparition of the King effect.

In the absence of this effect, the whole distribution is given by the equation (41). This may expressed as

$$R(S) = \mu + (1/\beta)\text{Ln}[S/(S-1)] \qquad (36)$$

which gives for large S

$$R(S) = \mu + (1/\beta)\,(1/S) \qquad (37)$$

One recovers the power law with a negligible correction. The exponent is 1 and this model cannot give other values for the exponent.

Maslov and Maslova (MM) (2006) also used the Bose-Einstein statistics for the distribution of words in a given language, but with different hypothesis from those of Yukalov and Sornette (2012) (YS), since the equivalence with the group of atoms following the Bose-Einstein statistics is different from of YS. For MM the words are the particles (not necessarily atoms, see below), the different frequencies of words are the energy levels (these are ranks in the YS version), and the number of words with the same frequency corresponds to population of one energy level.
First, they define a virtual language, which is deduced from the real language by suppressing the ambiguities of the texts written in this language. This is done principally in two ways: by replacing the pronouns by the person or the thing they represent and by clarifying the ellipsis. After this transformation, the length of a text is not well defined. More importantly, the new frequencies of words are higher than in the original language.
The main assumption of Maslov and Maslova is that the relation between the original frequencies $S_i$ and the new $\hat{S}_i$ are

$$\hat{S}_i = S_i(1 + \alpha S_i^\gamma) \qquad (38)$$

We recall that in the Bose-Einstein statistics a group of particles with a non-constant number, for example the photons, has a chemical potential equal to zero. In the linguistic analogy, a virtual language with an unknown length corresponds to a group of particles with a non-constant number. Accordingly, MM propose that the number of words (in the virtual language) with frequency $\hat{S}_i$ is

$$N_i = \frac{c}{\exp(\beta \hat{S}_i) - 1} \qquad (39)$$

It is the Bose-Einstein occupancy of particles with non-constant number (c is a constant).
The rank $R_j$ is given by

$$R_j = \sum_{i=1}^{j} N_i$$

And taking account of (38) and (39) one gets for small β

$$R_j = T \ln\left[\rho \frac{S_j^\gamma}{(1+\alpha S_j^\gamma)}\right] \qquad (40)$$

where $T = 1/\beta$ and $\rho = \exp(c/T)$.
For large $S_j$ one gets

$$\ln R = \text{Constant} - \gamma \ln S \qquad (41)$$

One recovers, for large enough S, the power law with exponent $\gamma$.
Finally, we mention that distributions based on the Bose Einstein statistics belong to type C.

## C. Human Behavior

In a very interesting article, Blasius and Tonjes (2009) show that the popularity or more precisely the frequency of the chess openings has a power law statistical distribution.
To explain this result they consider the tree of the game. Each step in the game may be viewed as the development of a tree as a random multiplicative process, defined by the branching ratio at each step $r_d = n_d / n_{d-1}$. In this definition of the ratio, d is the number of steps from the beginning of the game, and $n_d$ is the frequency of the game at this step d. The basic equation for $n_d$ is as follows

$$n_d = N \prod_{i=1}^{d} r_i \qquad (42)$$

where N is the total of games. The ratio is distributed according to the function

$$q(r) = (2/\pi)(1-r^2)^{0.5} \qquad (43)$$

The final result of the calculation is that the PDF is proportional to

$$P(S) \approx [\ln(N/S)]^{d-1}(N/S)^{1-\beta} \qquad (44)$$

where $\beta$ is a parameter less than 1. For n<N, the expression (44) may be approximated by a power law with exponent $\alpha$ dependent linearly of the depth d of the game

$$\alpha = (1-\beta) + (d-1)/\text{Ln}(N) \qquad (45)$$

The real-life data are in a very good agreement with the calculation.

Maslov (2009) made a very interesting remark about the analysis of Blasius and Tonjes (2009). He notes that the probability distribution of the ratio r given by (43) is an empirical observation, which was applied in the calculation. He suggests that the divergence of the function q(r) for r = 1 is a sign of the skill of the players, since the majority of players in the database are well qualified.

## **D. Geographic and economical models**

In the first part of this section, we discuss four models with explicit application to City Size Distributions (CSD). They are intended to explain, why in CSD the exponent of the CDF is exactly 1. We would like to remark here, that there are other models dealing with variety of behaviors in the CSD (Benguigui and Blumenfeld-Lieberthal, 2009). In the second part of this section, we will discuss two models based on the qualitative point of view of Zipf himself (1949).

We begin with the model of Zanette and Manrubia (1997), based on a random multiplicative model with diffusion. The model setup consists of sites on a 2D lattice, occupied by the same population equal to 1 at time t = 0. At each step, two changes may take place. First, the population S(x,t) of a site located at the vector x may change following two possibilities

$$S(\mathbf{x},t) \to (1-q)p^{-1} S(\mathbf{x},t) \qquad (46a)$$
$$S(\mathbf{x},t) \to q(1-p)^{-1} S(\mathbf{x},t) \qquad (46b)$$

where p is a probability (0 < p < 1) and q is a parameter between 0 and 0.5. Secondly, a given site transfers a fraction $\alpha S(x,t)$ of his population to the neighbor sites. The authors solve the model numerically on a 200 x 200 square lattice with different values of parameters p, q and α. The most interesting and surprising result is that the distribution is a power law with exponent 2 (for the PDF), whatever the values of the parameters. The authors try to justify this result even when the system is found in a steady state. Note that in this model the number of cities is constant, but not the total population.

In the model of Marsili and Zhang (1998), the setup includes a system of cities with $S_i$ inhabitants in city i. New inhabitants may come to the city (labeled i) with probability $W_A(S_i)$ or inhabitants may leave the city with probability $W_D(S_i)$. The new inhabitants come from a reservoir of potential inhabitants or they are born in the city I (the model distinguishes between the two possibilities). There is also a small probability p that a new city is created with a single inhabitant. The master equation giving the average number q of cities with S inhabitants is

$$\Delta q(S,t) = W_D(S+1)q(S+1,t) - W_D(S)q(S,t) + W_A(S-1)q(S-1) - W_A(S)q(S,t) + p \quad (47)$$

The authors look for a solution in the stationary state $\Delta q = 0$, giving explicit expressions for the probabilities $W_A$ and $W_D$. The authors chose

$$W_A(S) = (1-p)S^\alpha/n \quad \text{and} \quad W_D = S^\alpha/n \quad (48)$$

where n is the total population and $\alpha \geq 1$. If $\alpha=1$, the authors interpret that as an indication that the inhabitants decide to move independently. For $\alpha=2$, the agents interact by pair and $\alpha = 3$ they interact by triplet, etc.

For $\alpha = 1$, the PDF is

$$D(S) \sim (1/S)\exp(-pS) \quad (49)$$

However, for $\alpha >1$ they get

$$D(S) \sim \left(\frac{p}{1-p}\right)\frac{(1-p)^S}{S^\alpha} \quad (50)$$

For $S < S^* = 1/p$, the PDF behaves as a power law with exponent close to $\alpha$, since p is chosen to be very small (order of $10^{-3}$). For $\alpha = 1$, the distribution is of type C2, which is rather rare. For larger values of $\alpha$ (but $\alpha < 2$), the distribution is of type B.

For $\alpha > 2$, almost all the population is concentrated in a megacity, which is much larger than all other cities. The authors conclude that only the case $\alpha=2$ is relevant, since it corresponds to the Zipf law. It seems that the authors are not familiar with the megacity phenomenon (the King effect, discussed above) and they miss the opportunity to give it a valid explanation.

The next two models we will now describe are based on the analysis of Zipf concerning human behavior. He proposes to distinguish between two choices of residential location. One is the principle of *Diversification,* in which population tends to move to the immediate source of raw materials. Consequently, the population will be dispersed among many widely scattered settlements. The second possibility is the principle of *Unification,* according to which the materials move towards the population, thus resulting in concentration of the population in one big city.

The model of Semboloni and Leyvraz (2004) is based on migration of inhabitants from city to city. Some individuals locate in large cities, as they can sell their products more easily at higher profit. Others prefer small cities, because they can more easily enjoy the resources of the city. Thus, these are the two different strategies, unification for immigration in large cities and diversification for small cities.

There are T agents and the population of a city of size s is normalized $s = S/T$ and $\Delta s = 1/T$ is the size of one agent. With a probability p, an agent will chose the unification strategy and with a probability $1- p$ he will chose the diversification strategy. Once he chose his strategy, he will move to an existing city with s inhabitants. In the first case, his choice is given by a probability proportional to $(s + \Delta s)$ and he moves randomly to one of the cities with s inhabitants. In the second case (diversification), the choice is given by a probability proportional to $1/(s + \Delta s)$ and he moves to one of the existing cities with s inhabitants.

The authors provide the analytical solution and finally they get

$$D(S) \sim (1/s^2 - \alpha) \qquad (51)$$

One recovers a power law for the $s^2 < \alpha$ (when α is parameter of the order 1). In this model, the number of inhabitants is constant but not the number of cities.

Another model, which attempts to provide an explanation for the alleged rank-size distribution of city sizes was proposed by Brakman et. al (1999). Their model is based on the well-known paper on the core-periphery spatial structure of economy by Krugman (1991), which in turn relies on the imperfect competition model developed by Dixit and Stiglitz (1977). Krugman shows how a combination of low transportation costs and economies of scale[1] in an economy can lead to, what is commonly referred to as "positive feedback", essentially a concentration of firms in certain places, while other places become "the periphery". The setup of the Krugman's model includes two regions with two types of products (agricultural and manufacturing) produced by two types of workers, while the manufacturing is characterized by the mobility of workers and economies of scale. A set of equations is developed for the two regions, defining total income, price index and the nominal wage[2] in each region. Krugman defines four parameters, which can offset the balance between the regions and create a concentration of manufacturing workers in one rather than the other (divergence). These four parameters are: the share of income spent on manufacturing, the fraction of manufacturing workers in each region, elasticity of substitution[3] (associated with the preference for variety) and transport costs of moving goods between regions. The model demonstrates that the combination of low transportation costs and economies of scale will create a circular causation, such that a region with a slightly higher share of workers employed in manufacturing will attract more and more workers due to lower price index and higher real wages.

---

[1] Economies of scale are economic benefits, which arise with the increase in the size of firms or scale of production, generally associated with decreasing average cost of output.
[2] A nominal wage (as opposed to real wage) is the payment received by an employee not adjusted for inflation.

Brakman et al (1999) refer to the Zipf Law as "an empirical regularity in search of a theory…", although they recognize that it does not always hold in reality. Furthermore, they discuss the different values of the slope of the log-log representation of the rank-size function and suggest that the slope varies relative to the structural economic changes taking place in an economy. They support the idea proposed by Parr (1985), who suggested that over time a nation will display an n-shaped pattern of the slope (the exponent in the power law referred to as Zipf). Furthermore, they provide the rank-size plots for three periods in the history of Netherlands: pre-industrialization (1600), industrialization (1900) and post-industrialization (1990) and the corresponding values of 0.55, 1.03 and 0.72 for the power law exponent. We may remark that, upon visual inspection, none of the plots would be classified as Strong Power Law.

In Brakman et al (1999), the Krugman's model is extended, in order to produce a distribution similar to that postulated by Zipf, rather than formation of a single large city or a number of large cities, with the rest of settlements regarded as rural hinterland. The extension is introduced as "congestion costs", associated with negative externalities arising in large cities, such as traffic congestion, pollution and crime. The simulation starts out with 24 cities, while each city receives a random share of manufacturing labor. In order to simulate the Zipf distribution, the authors set the values of parameters for the three above-mentioned periods as follows:

1. The pre-industrialization is characterized by relatively low share of manufacturing workers in the total labor force, weak economies of scale and high transportation costs;

2. The industrialization is characterized by increased share of manufacturing workers, significant economies of scale and major decrease in transportation costs. At the same time, the model assumes that there are no the negative feedbacks.

3. In the post-industrialization stage, the negative feedback is introduced into the model, in the form of additional costs stemming from the number of firms, which are already in the city, causing large cities to grow slower than the smaller ones.

The resulting rank-size plots for the three periods derived from the simulation, produce values of the exponent that are very similar to those calculated from the real data. The authors emphasize the usefulness of the general equilibrium economic model in creating an explanatory background to the rank-size distribution.

---

[3] Elasticity of substitution essentially measures how easily can a consumer substitute one consumption good for another (change the ratio between the quantities of the two goods), as a response to a relative change in the ration of prices of those goods.

## E. Criticality

The concept of criticality was used widely in phase transitions theory in physics. It was also used to explain the power law associated with the frequency of the earthquakes (Gutemberg-Richter law).

The models of Chen et al. (1991) and of Christensen and Olami (1992) are both based on the original model of Burridge and Knoppof (1967). The general concept of the original model lies in division of space into blocks connected by springs. In the model of Chen et al, a shear stress is applied to opposite sides of the system in the shape of a square or a cube. When level of stress is increased, some of the springs connecting the blocks receive stress higher than their threshold and break, causing redistribution of the elastic forces. The broken spring is replaced by a new one with a new threshold chosen randomly. Augmenting the external stress, a chain reaction occurs and a pseudo stationary state is established. Since the system is dynamic, after some time (i.e. following the increase of the external stress) a new chain reaction occurs. The occurrence of the chain reaction is analogous to an avalanche in a sand pile. The system may reach a stationary state, when the local release of stress balances the global increase of the strain on the system. According to the authors, this stationary state model is a good approximation of the real-life earthquake mechanism.

The model is solved by numerical simulation in 40x40 square and a 20x20x20 cube and effectively the distribution of the energy released in an earthquake is a power law. In 2D the exponent b of the Gutemberg-Richter law is 0.4 and in 3D it is 0.6.

## MODELS OF EXPONENTIAL AND STRECHED EXPONENTIAL DISTRIBUTIONS

As we showed in our first paper, phenomena characterized by an exponential distribution are quite rare. Same can be said of the models constructed to reproduce this distribution.

One such model, based on the distribution of money among a group of agents, was proposed by Dragulescu and Yakovenko (2000). The agents trade a certain amount δm of their money, step by step. The total number of agents N is constant as well as the total amount of money, M. The change in the amount of money $m_i$ of agent i and that of agent j from time t to time t + 1 is given by

$$m_i(t+1) + m_j(t+1) = m_i(t) + m_j(t) \qquad (52)$$

when $m_i(t) \geq 0$ for all i and all t. In other words, the model does not allow creation of debt.
The distribution of money among agents P(m) is the Boltzmann- Gibbs distribution

$$P(m) = C \exp[-(M/N)m] \qquad (53)$$

To get this result, a complete analogy is made with ideal gas. The atoms are perceived as agents, the total energy as total amount of money. The energy levels are given by a discretization of the total amount of money in amounts $m_b$ and the population of the level b is given by the number of agents $N_b$ with energy between $m_b$ and $m_b + \Delta m$. The distribution is given by maximizing the entropy of the system, as explained in statistical mechanics textbooks. To complete the analogy, a temperature T is defined as $1/T = M/N$. The validity of the expression (53) was verified by computer simulations, changing the amount of money exchanged δm from constant to random. Dragulescu and Yakovenko (2000) propose a more complex system, which gives the same Boltzmann-Gibbs distribution. They introduce into a system of agents, a firm which creates products, sells them, borrows money, returns it with interest, hires workers and pays wages. The model is solved by standard economic methods of maximizing profit, not analytically but by computer simulations.

In the section on the preferential attachment models, we saw the model of Krapivsky et al (2000), in which the probability for a new node to be connected to an already existing node is proportional to $S^\gamma$. They found that if $\gamma < 1$, the distribution is a stretched exponential one.

Frish and Sornette (1997) develop a very general model of stretched exponential distribution. They consider a product S of n independent random variables, $S = x_1 x_2 x_3 \ldots x_n$ and suppose certain properties for the pdf of one variable $x_i$, $p(x)$. In particular, they suppose that the function $f(x)$ is defined as $f(x) = \ln[p(e^x)] + x$ and has the following properties:
1. $f(x) \to \infty$ such that $f(x)$ may be normalized as a probability
2. the second derivative $f''(x) > 0$
3. $\lim[x^2 f''(x)] \to \infty$

If the independent PDF $p(x)$ is Gaussian, the authors show that the PDF of the variable S is a stretched exponential for large S.

## MODELS OF TYPE A1 DISTRIBUTION

In the above section about the power law models, we described models, which give power law and double Pareto behavior. In fact, the double Pareto distributions belong to type A1 distribution. We have already discussed them above, because they are variants of the random multiplicative process.

Clauset and Erwin (2008) proposed a model of multiplicative process with application to the distribution of body size of terrestrial mammals. The model considers the evolution of sizes of mammals (time step by time step) within a randomly selected population. At each step, an ancestor species creates a new species, the descendent species and the size $S_D$ of the descendant is related to the size $S_A$ of the ancestor by

$$S_D = AS_A \qquad (54)$$

where (same as in equation (1)) A is the stochastic variable. The difference between (54) and (1) is that $S_D$ and $S_A$ do not belong to the same species. Two other rules are added to the model: first, there is minimum size ratio $[S_D/S_A]_{min}$, which is the largest possible decrease in size; second, there is a probability p(S) that a species disappears. This probability is an increasing function of S. The authors tried two cases for p(S): a power law or a logarithmic function. The distribution U(A) of the variable A has a downward parabolic shape truncated at $A_{min} = [S_D/S_A]_{min}$, such that $U(A < A_{min}) = 0$.

The model is solved by numerical simulation and effectively the PDF has a maximum, when it goes to zero for large sizes. It is a type A1 distribution. The comparison with the data is good, but in the real data, the PDF has some undulations, that the model cannot reproduce.

Maclachlan (2009) presents a model of distribution of firm sizes, bases on random division of an interval. The aim of the model is to find the market shares for a fixed number of firms. A firm has a market share equal to the number of consumers, buying its product. Each buyer is supposed to have a unique first preference for a product made by one firm. The author calls this preference "a variety" and assigns a number to each "variety". The consumers can then be arranged in the order according to the number of their "variety". The assumptions of the model are:

1. A buyer will chose the firm, which offers a product with the closest variety number to his personal preference.
2. The firm's manager will decide to differentiate his product from others, but he does this at random.

The model is solved by numerical simulation for 100 firms and the author presents their PDF, as well as the Rank-Size distributions. The results are typical of an A1 type with a maximum in the PDF and an infinite slope in the Rank-Size graph. However, the author shows that, although the PDF is reminiscent of a lognormal distribution, in fact it is different. The large size tail can be approximate by a power law with an exponent near 0.5.

Bhattacharyya et al (2007) solve a model of trading between agents exchanging money. The total amount of money is constant as well the number of agents. It is similar to the model by Dragulescu and Yakovenko (2000) which is discussed above. When two agents (labeled i and j) make a transaction, they exchange some money, but at the same time they save a part of their money λm, where λ is a stochastic variable.

The equation of the model for the money $m_i(t)$ of the agent i at time t+1 is

$$m_i(t + 1) = \lambda_i m_i(t) + \varepsilon[(1 − \lambda_i)m_i(t) + (1- \lambda_j)m_j(t)] \qquad (55)$$

where $\lambda_i$ and $\lambda_j$ are stochastic variables and ε is a parameter smaller than 1, constant or variable with t. An analogous equation defines the amount of money $m_j(t+1)$ for agent j. To solve the problem and find the distribution of the agents' money the authors multiply the two equations giving $m_i(t+1)$ and $m_j(t+1)$ and apply a kind of mean field approximation. They replace the quadratic quantities $m_i^2$, $m_j^2$ and $m_i m_j$ by a mean quantity $m^2$. Finally, they get the following equation for $m^2$,

$$m^2(t+1) = \eta(t)m^2(t) \qquad (56)$$

where the stochastic variable η(t) is a function of $\lambda_i$, $\lambda_j$ and ε. The solution of (56) gives the PDF of m as

$$P(m) = (1/m^2)\exp[-(\log(m^2))^2] \qquad (57)$$

This is a lognormal distribution for $m^2$ and an A1 distribution for m.

The distribution of the world languages was the subject of several papers and we chose to discuss the model, which displays a qualitative agreement with the real data, that of Stauffer et all (2007). The model describes a language as composed of F features (for example, grammar rules) and each feature has Q characteristics. The authors take the simplest case using Q = 2. Therefore, a language is defined by a string of 16 bits (features) with values of 1 and 0. The authors tried longer strings, but the results are qualitatively similar. Two languages are different, if they differ by at least by one bit.

The simulation begins with a population of N persons and at each step a person gives birth to one child or a person dies with a probability proportional to the total number of persons N(t). At birth, the child adopts the mother tongue, but there is a possibility to switch to a different language by switching one bit of the mother language. This happens with probability p. The choice of the bit to be switched is made with probability q. Finally speakers of a small language (the number of people speaking this language is small) switch to a larger language with probability r, which is a quadratic function of the fraction $x_i$ of the size of language i.

Depending of the initial conditions, there can be two outcomes at equilibrium: either one language is dominant (more than 3/4 of the total population speak this language) or fragmentation into all the possible languages, which are 65,536 in this model of 16 bits.

Thus, the model needs additional hypothesis in order to be analogous with reality. The first additional hypothesis is "irreversibility": when at a given step a language is changed by switching one bit, this is possible only if bit 0 is changed to bit 1, but not from bit 1 to bit 0. The second hypothesis is the introduction of noise by multiplying each language by a parameter s at each iteration. This parameter is obtained by multiplying ten random numbers taken in the interval [0,1]. Finally, the authors do not look for equilibrium, but only for situations in which the simulated distribution is similar to that of actual languages. Best results were obtained with only the second modification (noise) and the simulated distribution is qualitatively similar to a lognormal distribution.

It is clear, that one cannot expect a good agreement applying such a crude model. This is also the opinion of the authors. At the end of their paper, since they suggest that at equilibrium the largest language (today it is Chinese) will dominate everything, the authors write: *"But since it is only a computer model, it does not prove that in the future everybody will speak Mandarin Chinese and its mutants"*.

CONCLUSION

In this paper, we have presented a selection of models that simulate the distributions of phenomena from very different disciplines. These models were constructed in an attempt to explain the mechanisms, underlying the observed regularities in the distributions of phenomena in natural and social sciences. In most cases, the simulated distributions are compared to the original real-life distribution to validate the accuracy of the model. The literature on the subject is enormous and we were able to detail only a small fraction of the existing models. Some of the models are extremely complex and their construction involves defining multiple parameters, assumptions and constraints, as well as understanding of the processes they imitate. In this study, we chose to analyze some of the most general and often used models, from different disciplines and spheres of knowledge, in which the type of distribution they produce plays an important role. The type of models, which we discuss in this paper, can be grouped into a few broad categories. The first group, which we may refer to as the "physicists group" since the majority of such models were proposed by physicists, includes simple models with a relatively small number of parameters. One of their characteristics is that one model may be applied in different fields of research. We saw numerous examples above and the most known models of this group are the random multiplicative process and the preferential attachment models. Their weak point is that they are simplified models to describe a complicated reality.

The models of the second group are built with the explicit goal to be applied to a specific field, although it is possible to apply them to other situations. In this group, one finds models in economic geography (for example, growth of cities), in biology, in linguistics and other fields. These models are, in general, very complicated since they include a large numbers of parameters. This is the reason why we have presented only a few examples. The logic in building such models is to reproduce the complex reality as closely as possible. Admittedly, these models are generally successful in producing the shape of the distribution they studied. However, despite the success, the weakness of such models is precisely the large number of parameters. It is not possible to understand, qualitatively, the role of each parameter. It is like a black box, in which the parameters are introduced and the output of the box gives the distribution. The SIMPOP model (Sanders et all 1997) is a good example of such a situation.

One can ask the question, whether it is possible to construct models, which will be simple and nevertheless not too far from the real situations. We propose to look for the "lump method" in which a relatively small number of parameters are lumped from other more "real" parameters. Certainly, it is not a simple task but may be worth to try.

As we noted in our previous paper on the classification of distributions, there is a general tendency to assume that the distribution is a regular one, and many researchers are attracted by the idea that phenomena from different spheres of life can all be approximated by a universal mathematical expression. This regularity of the data is very often assumed to follow a power law, although in many cases, the empirical evidence is rather ambiguous. For a reason that remains to be understood, many researchers are particularly drawn to the power law and it is often being referred to as a self-evident truth. Possibly, this is also due to its convenient linear representation in log-log for Rank – Size, which is very often used as a criteria for the existence of power law (see the article of Maslov and Maslova (2006) for their doubts about the quality of a fit with the Rank-Size plot). The alleged universality of the power law is often supported by authors due to their reliance on only one type of graphic representations (mainly, the Rank-Size), which can be very misleading (see Appendix). In this study, we also review a number of models, which produce exponential distribution, lognormal distribution and other types of distributions that can be classified as type A1. However, these are not as popular.

In our previous study, we have shown that the true power law is not nearly as common as we have been led to think. The survey included 89 studies of the different phenomena, with only about one third of the complete distributions well described by a power law. In the present survey we showed that models giving a Strong Power Law, i.e. a power law for the complete distribution, are a minority as well.

Due to this power law assumption, many authors look only at a part of the modelled data (we refer to this as "Partial Power Law"). This generally necessitates a cut-off value, thus omitting the smaller or the larger items. In some of the models, a constraint is introduced, defining one of the parameters or holding the number of items constant. For example, in the Gibrat model, the power law can be obtained only for cases where parameter B in the equation (1) is set to equal zero (see table 1 above). For the other cases, the model produces only a partial power law or a double Pareto distribution. The double Pareto distribution is another such example, where a power law does not approximate the distribution with a single exponent, but by two different power laws, holding for the smaller and the larger items in the distribution. Although some authors claim that they are only looking for models giving the power law in the tail (see the article of Frisch and Sornette (1997) for justification of this procedure), in fact they might be missing significant and interesting points and, the type of distribution attributed to the data may be unconnected to the processes, suggested by the model. For example, the PDF of the lognormal distribution displays a maximum for the small values. If this range of small items is omitted from the data, it is easy to mistake the lognormal distribution for a power law, which likely is the case with at least some of the suggested models.

Even in the "True Power Law" cases, models giving the Zipf exponent of 1 (or rather -1), are rare, so that it may be considered an exception rather than the rule. It may be concluded, that the Zipf's Law holds mainly for special real data cases and models. We would like to suggest that the field could benefit from models, that are not a priori determined to find the power law, but first attempt to recognize the distribution, which best described the data, while keeping an open mind. This can hopefully, allow for more interesting and surprising results.

Finally, based on the discussion presented above, we may suggest that the use of the type of distribution as criterion for validation of the model may be insufficient. Clearly, very different models (simple as well as complex ones) can produce distributions, which are very similar or at least, appear to be similar. This is especially true when the authors intend to rely on the presence of the power law in order to demonstrate how successful their model is in replicating the reality. Even when more than one representation is used to examine the distribution, the correct recognition of the type of distribution may be difficult, due to certain assumptions and constraints introduced into the model (such as minimal or maximal thresholds or keeping the number of items constant throughout the "growth process"). We suggest that additional methods should be developed in order to test the validity and the accuracy of a specific model.

# APPENDIX

## Types of distributions in three representations

| Type | Rank Size | Cumulative | PDF |
|---|---|---|---|
| A1 | 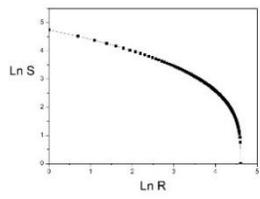 | 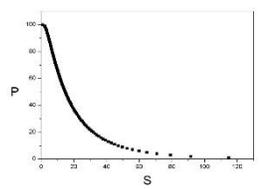 | 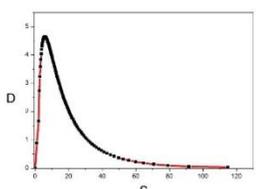 |
| A2 | 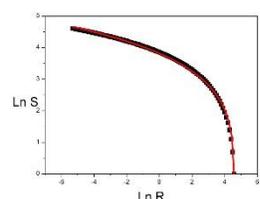 | 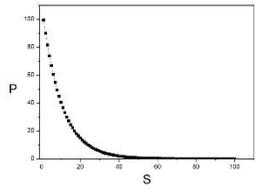 | 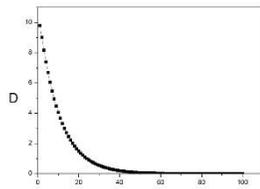 |
| A3 | 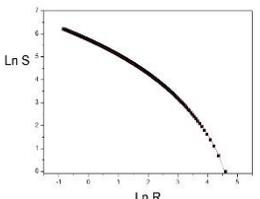 | 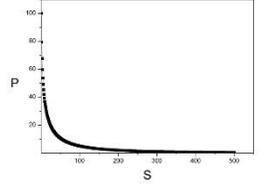 | 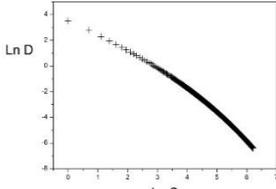 |
| B | 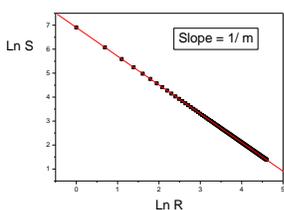 | 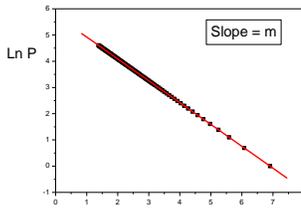 | 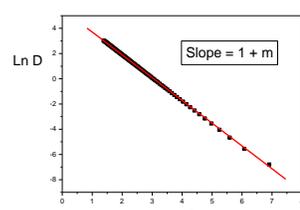 |
| **C1** | 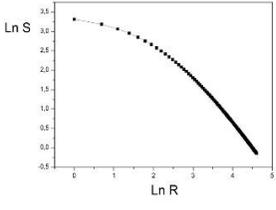 | 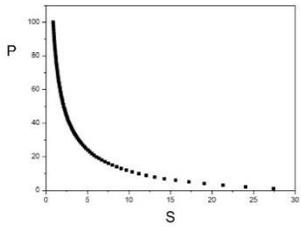 | 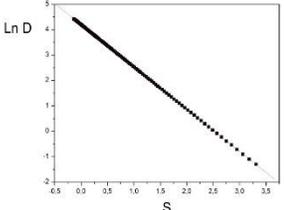 |
| **C2** | 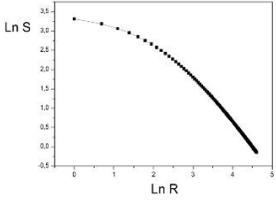 | 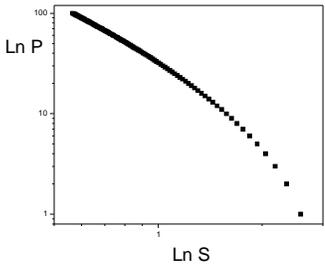 | 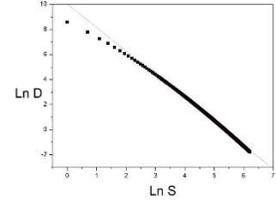 |
| **D** | 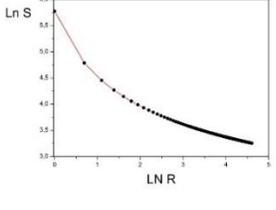 | 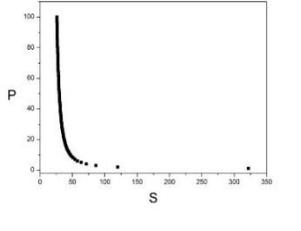 | 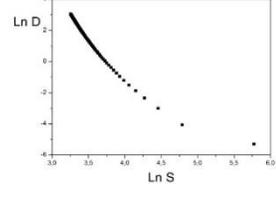 |
| **E** | Cases of non uniforn distributions. Example: species body sizes (Clauset and Erwin, 2008) 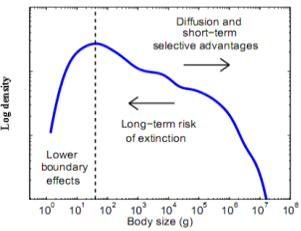 | | |

# REFERENCES

Barabasi A.L. and Albert R.(1999) *"Emergence of Scaling in Random Networks"* Science 286, 509-512.

Basseti B. Zarei M. Lagomarsino M.C. and Bianconi G.(2010) *"Statistical Mechanics of the Chinese Restaurant Process: lack of self-averaging, anomalous finite-size effects and condensation"* Phys.Rev.E **80.** 066118.

Bhattacharyya P., Chatterjee A. and Chakrabarti B.K. (2007) *"A common mode of origin of power laws in models of market and earthquake"* Physica A **381,** 377-382.

Benguigui L. and Marinov M. *"A Classification of Natural and Social Distributions Part One: the Descriptions"* arXiv:1507.03408v1.

Benguigui L. and Blumenfeld-Lieberthal E. (2007) *"A dynamic model for city size distribution beyond Zip's law"* Physica A **384,** 613-627.

Benguigui L. and Blumenfeld-Lieberthal E. (2009) *"The temporal evolution of the city size distribution"* Physica A **388,** 1187-1195.

Blank A. and Solomon S. (2000) *"Power laws in cities population, financial markets and internet sites (scaling in systems with a variable number of components)"* Physica A **287,** 279-288.

Blasius B. and Tonjes R. (2009) *"Zipf law in the popularity distribution of chess openings"* Phys.Rev.Lett. **103,** 218701.

Brakman S. Garretsen H. Van Marrewijk C. and van den Berg M. (1999) *"The return of Zipf: towards a further understanding of the rank-size distribution"* J.Reg.Science **39,** 183-213.

Burridge R. and Knoff (1967) *"Model and theoretical seismicity"* Bull. Seismological Society of America **57,** 341- 371.

Champernowne D.G. (1953) *"A Model of Income Distribution"* The Economic Journal, **250**, 318-353.

Chen K. Bak P. and Obukhov (1991) *"Self-organized criticality in a crack-propagation model of earthquakes:* Phys.Rev.Lett. **43,** 625-629.

Christensen K. and Olami Z. (1992) *"Variation of the Gutenberg-Richter b Values and Nontrivial Temporal Correlation in a Spring-Block Model for Earthquakes"* J.Geo.Research **47,** 8729-8735.

Clauset A. and Erwing D.H. (2008) "*The Evolution and Distribution of Species Body Size*" Science **321,** 399-401.

Clauset A. (2011) "*Inference, Models and Simulation for Complex Systems*" CSCI 7000/4830 (Lecture 14).

Corominas-Murtra B. and Solé R. (2010) "*Universality of Zipf's Law*" Phys.Rev.E **82**, 011102.

Dragulescu A. and Yakovenko V.M. (2000) "*Statistical mechanics of money*" Eur.Phys.J. B **17,** 723-729.

Frisch U. and Sornette D. (1997) "*Extreme Deviations and Applications"* J.Phys. **7,** 1151-1171.

Gabaix X. (1999) "*Zipf's law for cities: an explanation"* Quart. J. Econ. **114**, 739–67.

Gabaix X. (2010 "*Power Laws in Economics and Finance"* Annu. Rev. Econ. **1**, 255–293.

Kesten H. (1973). "*Random difference equations and renewal theory for products of random matrices",* Acta Math. **131**:207–48.

Krapivsky P.L. Redner S. and Leyvraz F. (2000) "*Connectivity of Growing Random Networks"* Phys.Rev.Lett. **85**, 4629-4632.

Krugman P. (1991) "*Increasing Returns and Economic Geography"* J.Political Economy **93,** 483-499.

Levy M. and Solomom S. (1996) *"Dynamical Explanation for the Emergence of Power Law in a Stock Market"* Int.J.Mod.Physics **7,** 65-72.

Maclachlan F.C. (2009 *"Random Division and the Size Distribution of Business Firms"* Complex Systems **18,** 227-236.

Malcai O., Biham O. and Solomom S. (1999) *"Power-Law Distributions and Lévy-Stable Intermittent Fluctuations in Stochastic Systems of Many Autocatalytic Elements"* Phys.Rev,E **60,** 1299-303.

Malevergne Y. Saichev A. and Sornette D. (2008) "*Zipf's Law for Firms: Relevance of Birth and Death Processes"*. Available at SSRN: http://ssrn.com/abstract=1083962 or http://dx.doi.org/10.2139/ssrn.1083962.

Marcili M. and Zhang Y.C. (1998) "*Interacting Individuals Leading to Zipf's Law"* Phys.Rev.Lett. **80**, 2741-2745.

Maruvka Y.E. Kessler D.A. and Shnerb N.M. (2011) *"Birth-Death-Mutation Process: a New Paradigm for Fat Tailed Distributions"* PLoS One 6. E26480.

Maslov S. (2009) "*Power laws in chess*" Physics **2,** 97-99.

Maslov V.P. and Maslova T.V. (2006) "*On Zipf's Law and Rank Distributions in Linguistics and Semiotics*" Mathematical Notes **80**, 679-691.

Mitzenmacher, M. (2003) "*A brief history of generative models for power law and lognormal distributions*" Internet Math. **1**, 226–51.

Par J.B. and Suzuki K. (1973) "*Settlement Populations and the Lognormal Distribution*" Urban Studies **10***,* 335.

Pumain D. (2012) *"Une Théorie Géographique pour la Loi de Zipf"* Région et Développement **36**, 31-54.

Reed W.J. (2002) *"On the Rank-Size Distribution for Human Settlements"* J.Reg.Science **42,** 1-17.

Reed W.J. (2003) *"The Pareto of incomes-an explanation and an extension"* Physica A **319,** 469-486*.*

Sanders L Pumain D. Mathian H. Guérin-Pace F. and Bura S.(1997) *"SIMPOP: a multiagent system for the study of urbanism"* Environment and Planning B **24,** 287-305.

Sembolini F. and Leyvraz F. (2005) "*Size and resources driven migration resulting in a power law distribution of cities"* Physica A **352,** 612-628.

Schulze C., Stauffer D, and Wichmann S. (2007) "*Birth, survival and death of languages by Monte-Carlo simulation",* arXiv 0704.0691.

Simon H. (1955) "*On a Class of Skew Distribution Functions"* Biometrika **42,** 425**-**440.

Sornette D. and Cont R. (1997) *"Convergent Multiplicative Processed Repelled from Zero: Power Laws and Truncated Power Law"* J.Phys. **7,** 431-444.

Takayasu  H. .Sato H.A. and Takayasu M. (1997) *"Stable Infinite Variance Fluctuations in Randomly Applied Langevin Systems"* Phys.Rev.Lett. **79,** 966-969.

Yukalov V.I. and Sornette D. (2012) *"Statistics outliers and dragon-kings as Bose condensed droplets"* Eur.Phys.J. Special Topics **205,** 53-64.

Zanette D. and Montemuro M.A. (2005) *"Dynamics of text generations with realistic Zipf distribution",* J. Quant. Linguistics **12,** 29-40.

Zanette D. and Manrubia S.C. (1997) *"Role of Intermittency in Urban Development: A Model Large-Scale City Formation"* Phys.Rev.Lett., **79,** 523-528.

Zipf G.K. (1949), *"Human Behavior and the Principle of Least Effort",* AddisonWesley.